\documentstyle[preprint,epsfig,aps]{revtex}

\tightenlines
\draft

\begin{document}

\title{Probability Distribution of the Shortest Path on the Percolation Cluster,
       its Backbone and Skeleton}

\author{Markus~Porto$^{1,2}$, Shlomo~Havlin$^2$, H.~Eduardo~Roman$^{3}$,
        and Armin~Bunde$^1$}

\address{$^1$Institut f\"ur Theoretische Physik III,
         Justus-Liebig-Universit\"at~Giessen,
         Heinrich-Buff-Ring~16, 35392~Giessen, Germany}
\address{$^2$Minerva Center and Department of Physics,
         Bar-Ilan~University,
         52900~Ramat-Gan, Israel}
\address{$^3$Dipartimento di Fisica,
         Universit\`a di Milano,
         Via~Celoria~16, 20133~Milano, Italy}

\date{received \today}

\maketitle

\begin{abstract}
We consider the mean distribution functions $\Phi(r|\ell)$, $\Phi_{\rm
B}(r|\ell)$, and $\Phi_{\rm S}(r|\ell)$, giving the probability that two sites
on the incipient percolation cluster, on its backbone and on its skeleton,
respectively, connected by a shortest path of length $\ell$ are separated by an
Euclidean distance $r$. Following a scaling argument due to de~Gennes for
self-avoiding walks, we derive analytical expressions for the exponents $g_1 =
d_f + d_{\rm min} - d$ and $g_1^{\rm B} = g_1^{\rm S} = 3 d_{\rm min} - d$,
which determine the scaling behavior of the distribution functions in the limit
$x \equiv r/\ell^{\tilde{\nu}} \ll 1$, i.e.\ $\Phi(r|\ell) \propto
\ell^{-\tilde{\nu}d} \; x^{g_1}$, $\Phi_{\rm B}(r|\ell) \propto
\ell^{-\tilde{\nu}d} \; x^{g_1^{\rm B}}$, and $\Phi_{\rm S}(r|\ell) \propto
\ell^{-\tilde{\nu}d} \; x^{g_1^{\rm S}}$, with $\tilde{\nu} \equiv 1/d_{\rm
min}$, where $d_f$ and $d_{\rm min}$ are the fractal dimensions of the
percolation cluster and the shortest path, respectively. The theoretical
predictions for $g_1$, $g_1^{\rm B}$ and $g_1^{\rm S}$ are in very good
agreement with our numerical results.
\end{abstract}

\bigskip
\pacs{PACS numbers: 05.20.$-$y, 64.60.$-$i, 05.40.$+$j}
\newpage
\narrowtext

Percolation constitutes a useful model for a variety of disordered systems in
many fields of science displaying both structural disorder and self-similarity
(i.e.\ fractal behavior) within some range of length scales
\cite{Bunde/Havlin:1996+Sahimi:1993+Stauffer/Aharony:1992}. In many
circumstances, the knowledge of the internal structure of percolation clusters
is required, as for instance in the study of transport processes near the
percolation threshold $p_{\rm c}$, where the complex topology of the available
conducting paths play a crucial role
\cite{Alexander/Orbach:1982,Havlin/Ben-Avraham:1987,Bunde/Havlin/Roman:1990,Nakayama/Yakubo/Orbach:1994}.

It is known that at the percolation threshold $p_{\rm c}$, the incipient
infinite cluster displays fractal behavior over all length scales, i.e.\ its
mass $s$ contained within a distance $r$ from a given cluster site chosen as
the origin, averaged over many origins, scales as $s \propto r^{d_f}$, where
$d_f = 91/48$ in two dimensions, $d_f = 2.524 \pm 0.008$ in three dimensions,
and $d_f = 4$ above the critical dimension, i.e.\ when $d \ge d_{\rm c} = 6$
\cite{Bunde/Havlin:1996+Sahimi:1993+Stauffer/Aharony:1992}. A second, useful
metric is the ``chemical'' distance $\ell$ between two cluster sites
\cite{Havlin/Ben-Avraham:1987}, defined as the length of the shortest path
connecting them. It is found that the mean distance $r$ between two cluster
sites, averaged over many pairs of sites, behaves as a function of $\ell$ as $r
\propto \ell^{1/d_{\rm min}}$, where $d_{\rm min} = 1.130 \pm 0.004$ in $d=2$
\cite{Herrmann/Stanley:1988}, $d_{\rm min} = 1.374 \pm 0.004$ in $d=3$
\cite{Grassberger:1992}, and $d_{\rm min} = 2$ when $d\ge d_{\rm c}$, is the
so-called fractal dimension of the shortest path. From the above scaling
relations follow that in ``chemical'' space, the mass of the cluster scales
with distance $\ell$ as $s \propto \ell^{d_{\ell}}$, where $d_{\ell} =
d_f/d_{\rm min}$, with $d_{\ell}=2$ when $d \ge d_{\rm c}$
\cite{Havlin/Ben-Avraham:1987}.

The incipient infinite cluster exhibits a variety of substructures that are
self-similar as well
\cite{Bunde/Havlin:1996+Sahimi:1993+Stauffer/Aharony:1992}. A prominent example
is the backbone of the cluster, defined as the subset of cluster sites that can
carry a current when a potential difference is applied between two sites (see
\cite{Herrmann/Hong/Stanley:1984} and references therein). Thus, the structure
of the backbone alone determines the conductivity of the whole percolation
network between two sites. The structural and dynamical properties of the
backbone of the incipient cluster have been studied recently
\cite{Porto/Bunde/Havlin/Roman:1997}. A second cluster substructure, denoted as
the skeleton (a subset of the backbone, also called the ``elastic'' backbone)
is defined as the union of all shortest paths between the two cluster sites.

In this Rapid Communication, we extend our previous studies of the structural
properties of the incipient infinite cluster \cite{Roman:1995} and its backbone
\cite{Porto/Bunde/Havlin/Roman:1997} in two and three dimensions. We consider
the structural distribution function $\Phi(r|\ell)$ for the incipient infinite
cluster, where $\Phi(r|\ell) \; {\rm d}r$ is the probability that two cluster
sites connected by a shortest path of length $\ell$ are at Euclidean distance
between $r$ and $r+{\rm d}r$ from each other in space. The probability
distribution $\Phi(r|\ell)$ is normalized according to $\int r^{d-1} \;
\Phi(r|\ell) \; {\rm d}r = 1$, and is found to obey an scaling behavior with
the variable $x \equiv r/\ell^{\tilde{\nu}}$ of the form $\Phi(r|\ell) =
\ell^{-\tilde{\nu} d} \; f(x)$ (see
e.g.~\cite{Havlin/Ben-Avraham:1987,Roman:1995,Neumann/Havlin:1988}), where
$\tilde{\nu}\equiv 1/d_{\rm min}$. Here, we draw our attention to the limit $x
\ll 1$, where the scaling function $f(x)$ follows a simple power law, $f(x)
\propto x^{g_1}$, i.e.
\begin{equation}\label{eq:Phirl}
\Phi(r|\ell) \propto \frac{1}{\ell^{\tilde{\nu} d}} \;
\left(\frac{r}{\ell^{\tilde{\nu}}}\right)^{g_1} \; ,
\quad\mbox{for $r/\ell^{\tilde{\nu}} \ll 1$} {\;}.
\end{equation}

Similar scaling forms for the substructural distribution functions as a
function of $x \equiv r/\ell^{\tilde{\nu}}$, $\Phi_{\rm B}(r|\ell) =
\ell^{-\tilde{\nu} d} \; f_{\rm B}(x)$ for the backbone and $\Phi_{\rm
S}(r|\ell) = \ell^{-\tilde{\nu} d} \; f_{\rm S}(x)$ for the skeleton, are
expected \cite{Havlin/Ben-Avraham:1987,Porto/Bunde/Havlin/Roman:1997}. In the
case $x \ll 1$, the corresponding scaling functions, $f_{\rm B}(x)$ and $f_{\rm
S}(x)$, are found to behave as $f_{\rm B}(x) \propto x^{g_1^{\rm B}}$ and
$f_{\rm S}(x) \propto x^{g_1^{\rm S}}$, respectively, yielding
\begin{equation}\label{eq:PhirlBS}
\Phi_{\rm B}(r|\ell) \propto \frac{1}{\ell^{\tilde{\nu} d}} \;
\left(\frac{r}{\ell^{\tilde{\nu}}}\right)^{g_1^{\rm B}}
\quad\mbox{and}\quad
\Phi_{\rm S}(r|\ell) \propto \frac{1}{\ell^{\tilde{\nu} d}} \;
\left(\frac{r}{\ell^{\tilde{\nu}}}\right)^{g_1^{\rm S}} \; ,
\quad\mbox{for $r/\ell^{\tilde{\nu}} \ll 1$} {\;}.
\end{equation}

Numerical results (see Refs.~\cite{Porto/Bunde/Havlin/Roman:1997,Roman:1995}
and below) indicate that $g_1 < g_1^{\rm B} \cong g_1^{\rm S}$ in both two
and three dimensions. For $d \ge d_{\rm c}$, one expects the mean field (MF)
values $g_1 = g_1^{\rm B} = g_1^{\rm S} = 0$, since percolation clusters behave
similarly to simple random walks above the critical dimension $d_{\rm c}$
\cite{Roman:1995}.

\begin{figure}[h]
\epsfig{file=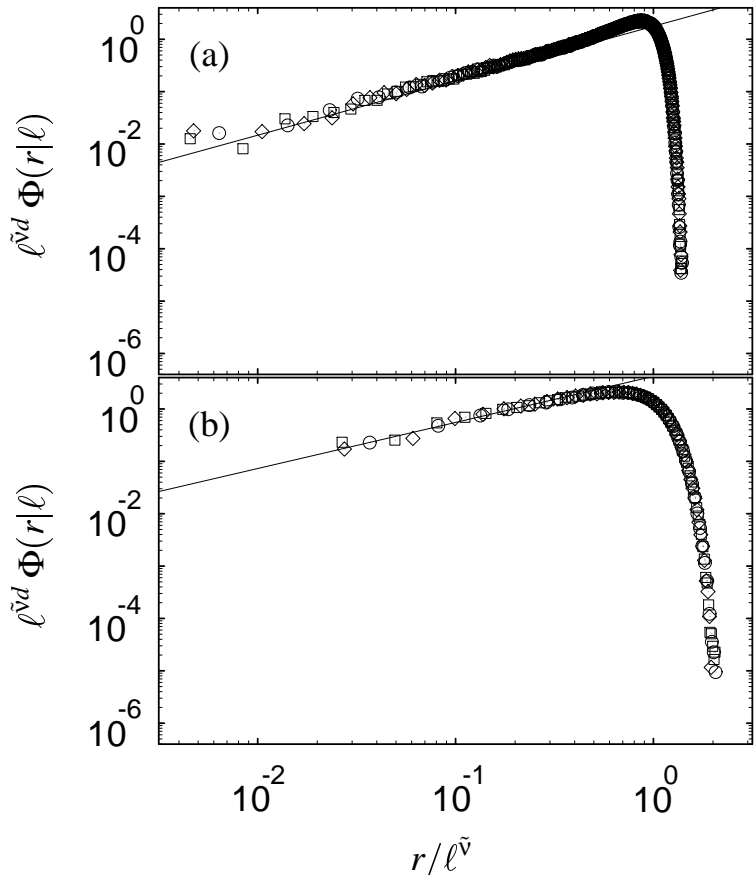,width=10cm,clip=,bbllx=50,bblly=110,bburx=560,bbury=710,angle=0}
\caption{Scaling plot of the probability distribution function
$\ell^{\tilde{\nu} d} \; \Phi(r|\ell)$ vs $r/\ell^{\tilde{\nu}}$ for the
incipient infinite cluster in the following cases: (a)~$d=2$, $\ell=1000$
(circle), $\ell=1400$ (diamond), and $\ell=1800$ (square), and, (b)~$d=3$,
$\ell=400$ (circle), $\ell=600$ (diamond), and $\ell=800$ (square). The plots
are based on averages over more than $10^5$ cluster configurations, for
clusters grown up to a maximum chemical distance $\ell_{\rm max} = 2000$ on a
square lattice ($d=2$) and $\ell_{\rm max} = 1000$ on a s.c.\ lattice ($d=3$).
The straight lines represent our fits for $\ell^{\tilde{\nu} d} \; \Phi(r|\ell)
= f(x)$ when $x \equiv r/\ell^{\tilde{\nu}} \ll 1$, and have the slopes
$g_1=1.04$ in (a), and $g_1=0.88$ in (b).}
\label{figure:Prl}
\end{figure}

\begin{figure}[ht]
\epsfig{file=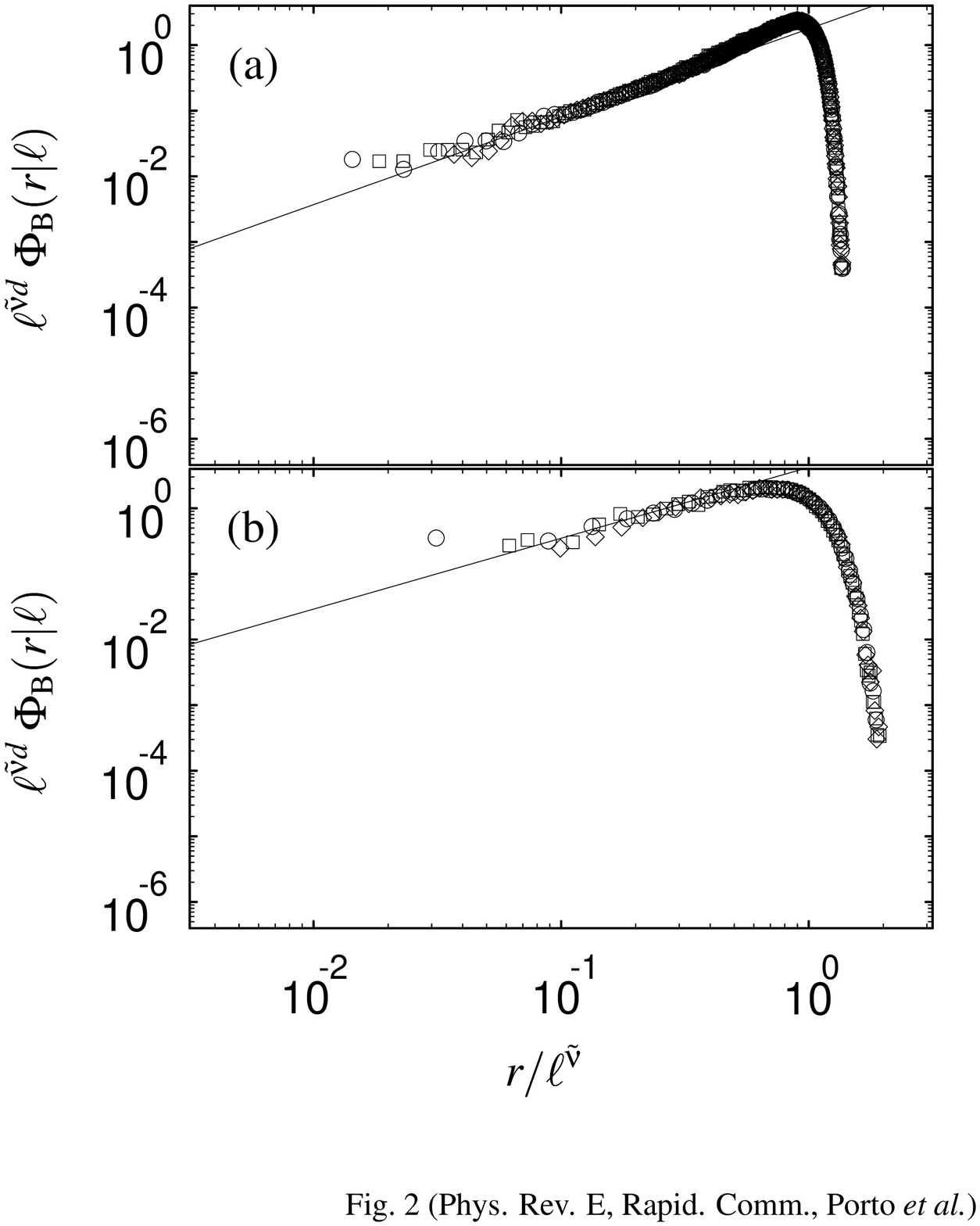,width=10cm,clip=,bbllx=50,bblly=110,bburx=560,bbury=710,angle=0}
\caption{Same as in Fig.~\ref{figure:Prl} for the probability distribution
function $\Phi_{\rm B}(r|\ell)$ of the backbone of the incipient cluster. The
straight lines have the slopes $g_1^{\rm B}=1.34$ in (a), and $g_1^{\rm
B}=1.08$ in (b).}
\label{figure:PrlB}
\end{figure}

\begin{figure}[ht]
\epsfig{file=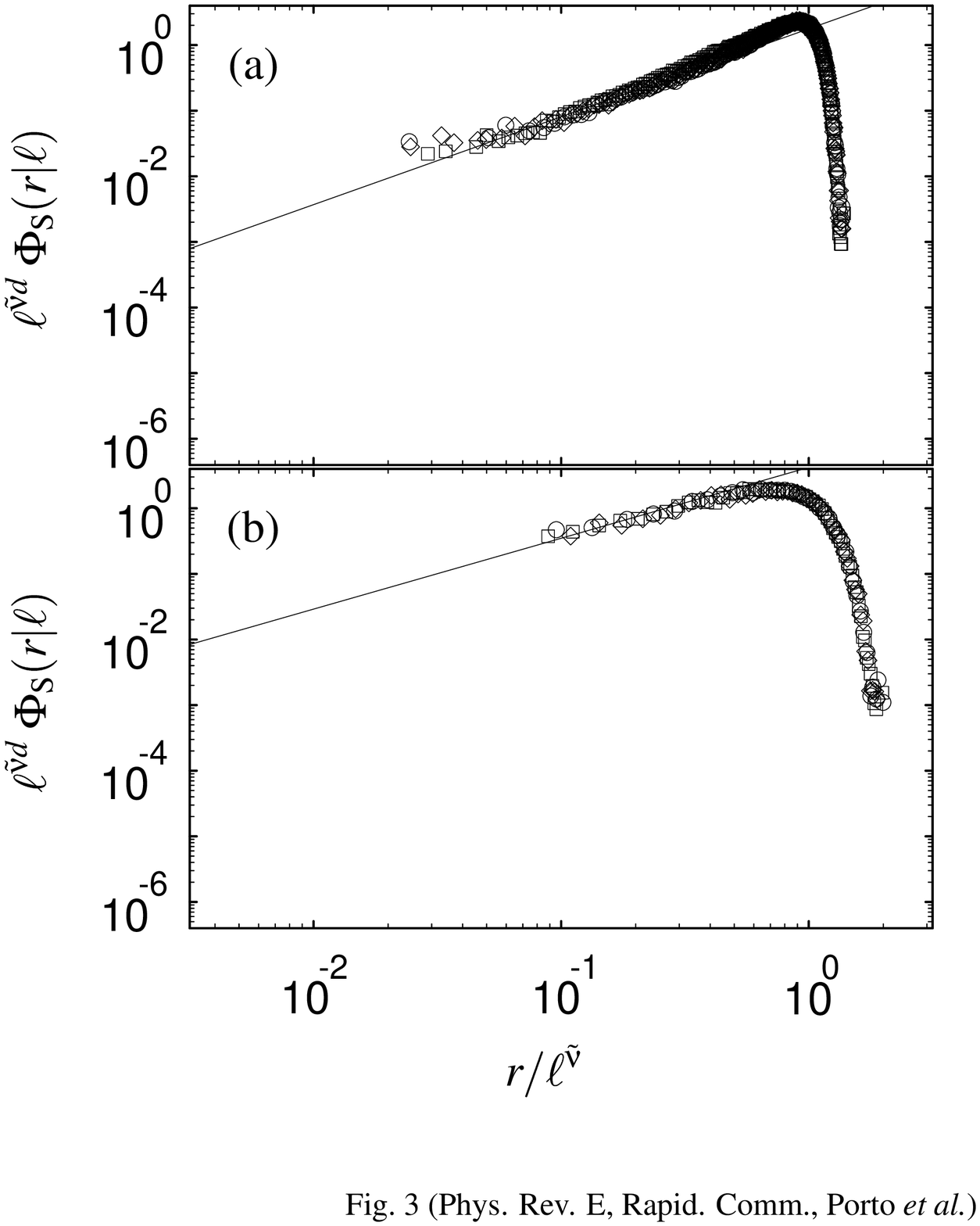,width=10cm,clip=,bbllx=50,bblly=110,bburx=560,bbury=710,angle=0}
\caption{Same as in Fig.~\ref{figure:Prl} for the probability distribution
function $\Phi_{\rm S}(r|\ell)$ of the skeleton of the incipient cluster. The
straight lines have the slopes $g_1^{\rm S}=1.34$ in (a), and $g_1^{\rm
S}=1.08$ in (b).}
\label{figure:PrlS}
\end{figure}

We first study the above defined distribution functions numerically, both in
two and three dimensions. To this end, we generate large percolation cluster at
$p_{\rm c}$ on square and s.c.\ lattices, respectively, using the well-known
Leath algorithm \cite{Leath:1976+Alexandrowicz:1980}. To identify the backbone
and skeleton of the cluster, we apply an improved version
\cite{Porto/Bunde/Havlin/Roman:1997} of the well-known burning algorithm
\cite{Herrmann/Hong/Stanley:1984}. We perform averages over more than $10^5$
clusters, which are grown until they reach a maximum of chemical shells
$\ell_{\rm max} = 2000$ in $d=2$ and $\ell_{\rm max} = 1000$ in $d=3$. The
results for $\Phi(r|\ell)$, $\Phi_{\rm B}(r|\ell)$, and $\Phi_{\rm S}(r|\ell)$
are shown in Figs.~\ref{figure:Prl}, \ref{figure:PrlB}, and \ref{figure:PrlS},
respectively. For the incipient infinite cluster we obtain $g_1 = 1.04\pm 0.05$
in $d=2$ and $g_1 = 0.88 \pm 0.05$ in $d=3$ (see also \cite{Roman:1995}). For
the backbone, we find $g_1^{\rm B} = 1.34 \pm 0.10$ in $d=2$ and $g_1^{\rm B} =
1.08 \pm 0.10$ in $d=3$ (see also \cite{Porto/Bunde/Havlin/Roman:1997}). In
addition, our results suggest that $\Phi_{\rm B}(r|\ell)$ and $\Phi_{\rm
S}(r|\ell)$ coincide, within the accuracy of the present data, and as a result,
the values of $g_1^{\rm S}$ for the skeleton are indistinguishable from those
of the backbone, i.e.\ $g_1^{\rm S} \cong g_1^{\rm B}$. These results are
summarized in Table I.

To estimate values for the exponents $g_1$, $g_1^{\rm B}$ and $g_1^{\rm S}$
analytically, we follow a method similar to the one discussed by de~Gennes
\cite{deGennes:1979} for determining the structure of self-avoiding walks (SAW)
of $N$ steps. The latter is described by the probability distribution function
$P_{\rm SAW}(r|N) = N^{-\nu d} \; f_{\rm SAW}(y)$, with $y \equiv r/N^{\nu}$,
where $P_{\rm SAW}(r|N) \; {\rm d}r$ gives the probability that the two end
points of a SAW of fixed length $N$ (i.e.\ the first and the $N+1$ monomers)
are at a distance between $r$ and $r+{\rm d}r$. Here, $\nu$ is the Flory
exponent, $\nu \cong (d+2)/3$ for $d \le 4$ and $\nu = \nu_{\rm MF} =
\frac{1}{2}$ for $d \ge 4$, and $f_{\rm SAW}(y)$ is the scaling function, with
$f_{\rm SAW}(y) \propto y^g$ when $y \ll 1$. For SAW defined on the lattice,
de~Gennes argues that the behavior of $f_{\rm SAW}(y)$ for $y \ll 1$ can be
obtained by considering the probability $P_{\rm SAW}(r\to 1|N)$ that a SAW of
$N\gg1$ steps returns close to its starting point (origin), which can be
written as
\begin{equation}\label{eq:PhirNSAWdeGennes}
P_{\rm SAW}(r \to 1|N) \propto
\frac{N_{\rm SAW}^{r\to 1}(N)}{N_{\rm SAW}(N)}
 {\;}, \quad\mbox{for $N \gg 1$} {\;},
\end{equation}
where $N_{\rm SAW}^{r \to 1}(N) \propto N^{-\nu d} \; \bar{z}^N$ is the number
of SAW of length $N$ returning close to the origin and $N_{\rm SAW}(N)
\propto N^{\gamma-1} \; \bar{z}^N$ is the total number of SAW of length $N$.
Here, $\bar{z}$ is the effective
coordination number of the lattice, and $\gamma$ is the enhancement exponent,
with $\gamma = \gamma_{\rm MF} = 1$ for $d \ge 4$.

As noted by de~Gennes \cite{deGennes:1979}, the enhancement factor
$N^{\gamma-1}$ occurs only in the denominator of the ratio $N_{\rm SAW}^{r \to
1}(N)/N_{\rm SAW}(N)$, but not in the numerator, indicating the ``difficulty''
for a SAW to return near to its starting point. Note that this missing
enhancement factor in the numerator can be viewed as corresponding to its
mean-field value, $N^{\gamma_{\rm MF}-1}\equiv1$, and one can write
equivalently
\begin{equation}\label{eq:PhirNSAW}
P_{\rm SAW}(r \to 1|N) \propto \frac{1}{N^{\nu d}}
\frac{N^{\gamma_{\rm MF}-1}}{N^{\gamma-1}}
 {\;}, \quad\mbox{for $N \gg 1$} {\;},
\end{equation}
corresponding to the behavior $f_{\rm SAW}(y) \propto y^g$, with $g =
(\gamma-1)/\nu$. This observation suggested us a procedure for describing the
structural function of the incipient percolation cluster and its substructures
analytically, in the case $r/\ell^{\tilde{\nu}} \ll 1$. We consider the
incipient percolation cluster first.

Let us generalize Eq.~(\ref{eq:PhirNSAW}) to percolation clusters by writing
the distribution function $\Phi(r|\ell)$, for a chemical distance $\ell \gg 1$
and Euclidean distance $r \to 1$, as
\begin{equation}\label{eq:Phi1l:Pi}
\Phi(r \to 1|\ell) \propto \frac{1}{\ell^{\tilde{\nu} d}}
\frac{\Pi_{\rm MF}(\ell)}{\Pi(\ell)}
 {\;}, \quad\mbox{for $\ell \gg 1$} {\;},
\end{equation}
where $\Pi(\ell)$ plays the role of the function $N^{\gamma-1}$ in
Eq.~(\ref{eq:PhirNSAW}), and $\Pi_{\rm MF}(\ell)$ denotes its mean field value.
Here we argue that, to a first approximation, $\Pi(\ell)$ is given by the
probability that the two chosen sites are on a cluster of chemical size $\ell$.
Therefore we relate $\Pi(\ell)$ to the probability distribution of cluster
sizes $s n(s)$, which is known to behave as $s n(s) \propto s^{-(\tau-1)}$,
with $\tau = 1+d/d_f$ for $d \le d_{\rm c}$, and $\tau_{\rm MF} = 5/2$
\cite{Bunde/Havlin:1996+Sahimi:1993+Stauffer/Aharony:1992}. Hence, $\Pi(\ell)$
is given by $\Pi(\ell) \propto s n(s) \; {\rm d}s/{\rm d}\ell$, and noting that
$s \propto \ell^{d_{\ell}}$, we find $\Pi(\ell) \propto \ell^{-d_{\ell}
(\tau-2)-1}$ for $d \le d_{\rm c}$, and $\Pi_{\rm MF}(\ell) \propto \ell^{-2}$.
Thus, Eq.~(\ref{eq:Phi1l:Pi}) becomes
\begin{equation}\label{eq:Phi1l:ell}
\Phi(r \to 1|\ell) \propto \frac{1}{\ell^{\tilde{\nu} d}}
\frac{\ell^{-2}}{\ell^{-d_{\ell} (\tau-2)-1}}
\propto \frac{1}{\ell^{\tilde{\nu} d}} \; \ell^{d_{\ell} (\tau-2)-1}
 {\;}, \quad\mbox{for $\ell \gg 1$} {\;}.
\end{equation}
Comparing this result with the one obtained from Eq.~(\ref{eq:Phirl}) in the
limit $r \to 1$, yields $-\tilde{\nu} g_1 = d_{\ell} (\tau-2)-1$, i.e.
\begin{equation}\label{eq:g1}
g_1 = d_f + d_{\rm min} - d {\;},
\end{equation}
which predicts $g_1 = 1.026 \pm 0.004$ for $d=2$ and $g_1 = 0.898 \pm 0.008$
for $d=3$. These theoretical values for $g_1$ are in very good agreement with
our numerical results (cf.\ Fig.~\ref{figure:Prl} and
Table~\ref{table:values}). Note that Eq.~(\ref{eq:g1}) yields by construction
$g_1 = 0$ for $d \ge d_{\rm c}$, as required.

The above argument can be applied straightforwardly to the backbone and the
skeleton of the incipient cluster, where now analogous equations to
Eq.~(\ref{eq:Phi1l:Pi}) can be written for $\Phi_{\rm B}(r \to 1|\ell)$ and
$\Phi_{\rm S}(r \to 1|\ell)$, with $\Pi(\ell)$ replaced by $\Pi_{\rm B}(\ell)$
and $\Pi_{\rm S}(\ell)$, respectively. In the case of the backbone, we argue
that $\Pi_{\rm B}(\ell) \propto n(s) \; {\rm d}s/{\rm d}\ell$, with $n(s)
\propto s^{-\tau}$, and $s \propto \ell^{d_{\ell}}$ as for the incipient
cluster. Note the absence of the factor $s$ in the expression for $\Pi_{\rm
B}(\ell)$, reflecting the fact that the backbone represents a subset of the
incipient cluster having a vanishing measure when $s\to\infty$ \cite{Note}.
Since the same argument applies to the skeleton, we have that $\Pi_{\rm
S}(\ell)\cong \Pi_{\rm B}(\ell)$, yielding
\begin{equation}\label{eq:Phi1lBS:Pi}
\Phi_{\rm S}(r \to 1|\ell)\cong \Phi_{\rm B}(r \to 1|\ell)
 {\;}, \quad\mbox{for $\ell \gg 1$} {\;},
\end{equation}
in agreement with the numerical results shown in Figs.~\ref{figure:PrlB} and
\ref{figure:PrlS}. In terms of $\Pi_{\rm B}(\ell)$, $\Phi_{\rm B}(r|\ell)$
in the limit $r \to 1$ is given by
\begin{equation}\label{eq:Phi1lBS:TildePi}
\Phi_{\rm B}(r \to 1|\ell) \propto \frac{1}{\ell^{\tilde{\nu} d}}
\frac{\Pi_{\rm B,MF}(\ell)}{\Pi_{\rm B}(\ell)}
 {\;}, \quad\mbox{for $\ell \gg 1$} {\;},
\end{equation}
and with $\Pi_{\rm B}(\ell) \propto \ell^{-d_{\ell} (\tau-1)-1}$
\begin{equation}\label{eq:Phi1lBS:ell}
\Phi_{\rm B}(r \to 1|\ell) \propto \frac{1}{\ell^{\tilde{\nu} d}}
\frac{\ell^{-4}}{\ell^{-d_{\ell} (\tau-1)-1}}
\propto \frac{1}{\ell^{\tilde{\nu} d}} \; \ell^{d_{\ell} (\tau-1)-3}
 {\;}, \quad\mbox{for $\ell \gg 1$} {\;}.
\end{equation}
Comparing this result with the scaling form for $\Phi_{\rm B}(r|\ell)$ given in
Eq.~(\ref{eq:PhirlBS}) in the limit $r \to 1$, yields $-\tilde{\nu} g_1^{\rm
B} = d_{\ell} (\tau-1)-3$, i.e.
\begin{equation}\label{eq:g1BS}
g_1^{\rm B} = 3 d_{\rm min} - d {\;},
\end{equation}
predicting $g_1^{\rm B} = 1.390 \pm 0.004$ in $d=2$ and $g_1^{\rm B} = 1.122
\pm 0.004$ in $d=3$, with $g_1^{\rm S} = g_1^{\rm B}$, in remarkable agreement
with the numerical results (cf.\ Figs.~\ref{figure:PrlB} and \ref{figure:PrlS},
and Table~\ref{table:values}). Note also that from Eqs.~(\ref{eq:g1BS}) and
(\ref{eq:Phi1lBS:Pi}) one obtains $g_1^{\rm B} = g_1^{\rm S} = 0$ for $d \ge
d_{\rm c}$, as expected.

In summary, we derive the analytical expressions $g_1 = d_f+d_{\rm min}-d$ and
$g_1^{\rm B} = g_1^{\rm S} = 3 d_{\rm min}-d$ describing the scaling behavior
of the structural distribution functions, $\Phi(r|\ell) \propto
\ell^{-\tilde{\nu}d} \; x^{g_1}$, $\Phi_{\rm B}(r|\ell) \propto
\ell^{-\tilde{\nu}d} \; x^{g_1^{\rm B}}$, and $\Phi_{\rm S}(r|\ell) \propto
\ell^{-\tilde{\nu}d} \; x^{g_1^{\rm S}}$, of the incipient percolation cluster,
its backbone and skeleton, respectively, at the critical concentration $p_{\rm
c}$ in the limit $x \equiv r/\ell^{\tilde{\nu}} \ll 1$. Here, $\tilde{\nu}
\equiv 1/d_{\rm min}$, and $d_f$ and $d_{\rm min}$ are the fractal dimensions
of the incipient percolation cluster and the shortest path, respectively. Note
that from the above expressions for the exponents $g_1$, $g_1^{\rm B}$, and
$g_1^{\rm S}$ follow that the corresponding distribution functions for $\ell
\gg 1$, in the limit $r \to 1$, scale as $\Phi(r \to 1|\ell) \propto
\ell^{-(d_{\ell}+1)}$ and $\Phi_{\rm B}(r \to 1|\ell) \cong \Phi_{\rm S}(r \to
1|\ell) \propto \ell^{-3}$, the latter being {\it independent\/} of the lattice
dimension $d$. We note that the result $\Phi(r \to 1|\ell) \propto
\ell^{-(d_{\ell}+1)}$ for $\ell \gg 1$, based on numerical simulations, was
also suggested for two other variants of percolation, invasion percolation with
as well as without trapping \cite{Schwarzer/Havlin/Bunde:1998}, and seems
therefore to be more general.

\bigskip
\bigskip

We acknowledge useful discussions with I.~Webman, J.~Dr\"ager and A.~Ordemann.
This work has been supported by the Minerva Center for the Physics of
Mesoscopics, Fractals and Neural Network; the German-Israeli Foundation; the
Alexan\-der von Hum\-boldt Foundation and the Deut\-sche
For\-schungs\-ge\-mein\-schaft.

\widetext

\begin{table}
\begin{tabular}{c||c|c||c|c||c}
 & \multicolumn{2}{c||}{$d=2$} & \multicolumn{2}{c||}{$d=3$} & $d=6$ \\
 & simulation & theory & simulation & theory & exact \\ \hline\hline
$g_1$         & $1.04\pm0.05$ & $1.026\pm0.004$ & $0.88\pm0.05$ & $0.898\pm0.008$ & $0$ \\
$g_1^{\rm B}$ & $1.34\pm0.10$ & $1.390\pm0.012$ & $1.08\pm0.10$ & $1.122\pm0.012$ & $0$ \\
$g_1^{\rm S}$ & $1.34\pm0.10$ & $1.390\pm0.012$ & $1.08\pm0.10$ & $1.122\pm0.012$ & $0$
\end{tabular}
\caption{Summary of the values for the exponents $g_1$, $g_1^{\rm B}$ and
$g_1^{\rm S}$ obtained from the numerical simulations and the analytic
expressions derived in the text.}
\label{table:values}
\end{table}

\begin{thebibliography}{99}
\bibitem{Note}
The quantity $s n(s) \propto s^{-(\tau-1)}$ gives the probability that a
randomly chosen site (or, equivalently, that a pair of randomly chosen sites)
is on a cluster of mass $s$, whereas $n(s) \propto s^{-\tau}$ gives the
probability that a randomly chosen cluster has a mass $s$.
\end{thebibliography}
\end{document}